\documentclass{ws-ijmpcs}

\begin{document}


%
%

\title{Testing antimatter gravity with muonium}

\author{KLAUS KIRCH}

\address{Institute for Particle Physics, ETH Zurich and\\
Paul Scherrer Institute, Switzerland.\\
klaus.kirch@psi.ch}

\author{KIM SIANG KHAW}

\address{Institute for Particle Physics, ETH Zurich, Switzerland.\\
khaw@phys.ethz.ch}

\maketitle



\begin{abstract}
The debate about how antimatter or different antimatter systems behave 
gravitationally will be ultimately decided by experiments
measuring directly the acceleration of various antimatter probes in the gravitational field of the Earth
or perhaps redshift effects in antimatter atoms caused by the annual variation of the Sun's gravitational
potential at the location of the Earth.
Muonium atoms may be used to probe the gravitational interaction of leptonic, second generation antimatter.
We discuss the progress of our work towards enabling such experiments with muonium.
\end{abstract}

\ccode{PACS numbers : 03.75.Dg, 04.80.Cc, 29.25.-t, 36.10.Ee, 41.75.Ak, 52.80.Dy, 67.25.D-, 76.75.+i}

\section{Introduction}	
The equivalence of the gravitational and the inertial mass of macroscopic test masses of
ordinary matter has been tested to very high precision, e.g. with a differential measurement
of Be and Ti masses using a highly sophisticated torsion balance~\cite{schlammi}.
Gravitational acceleration of ordinary matter atoms in the Earth gravitational
field has also been measured to high precision, e.g. using atom
interferometry methods~\cite{gravity1,gravity2}. Neutral kaon oscillations have been used to
set very tight constraints on differences in the gravitational interaction of their constituents,
see e.g.~\cite{cplear} and arguments have been extended using all 
neutral meson oscillations~\cite{karsh2}. While it may therefore be very improbable to find
some general 'antigravity', similar arguments as put forward already in~\cite{nieto} can
still be applied and the gravitational interaction may perhaps be much more complex and
allow for cancellations in some systems.
A first crude limit has been set on the antihydrogen gravitational
acceleration~\cite{alpha}, however,
neither decisive yet in its magnitude nor in its sign.
The precise measurement of antimatter gravity in various systems may serve as an 
important input for constructing theories of quantum gravity
and it could potentially provide insights to dark matter and dark energy
which remain mysterious until today.

Muonium (Mu) is the hydrogen like bound state of $\mu^{+}$ and $e^{-}$.
It has been studied extensively since the 1960s to test bound state QED,
to determine fundamental parameters and to search for exotic physics~\cite{hughes66,muqed1,muqed2}.
It can be reliably calculated within QED because it is purely leptonic and 
essentially free of hadronic effect. It can live relatively long, limited by
the muon life time ($\tau_{\mu}=2.2$~$\mu$s). 
Its 1S-2S transition
frequency~\cite{mu1s2s1,mu1s2s2} and the hyperfine splitting of the ground
state~\cite{muhfs} have been measured to high precision (4 ppb and 12 ppb).

In recent years, there is a renewed interest in physics with Mu atoms, triggered by new ideas
for much improved slow muon and muonium beams and by progress in laser technology.
Measuring Mu's gravitational interaction with ordinary matter would be complementary to
such measurements with antihydrogen and positronium and would be the first test for this
second generation leptonic system where the mass is dominated by the heavy $\mu^+$.

Two very different approaches for a Mu gravity experiment are being considered:

\begin{itemize}
	\item{Search for an annual modulation of the Mu(1S-2S) transition frequency~\cite{kars}}
	\item{Use a high quality Mu beam passing through a Mach Zehnder atom interferometer~\cite{mzinter1,mzinter2,kirch,kaplan}}
\end{itemize}

Our group is pursuing research into both directions and developing the necessary experimental prerequisites. Some technical challenges and the current situation of research and development will be sketched in the following sections.

\section{Experimental challenges}

\subsection{1S-2S spectroscopy}
First proposed for positronium 1S-2S spectroscopy~\cite{kars,ps1s2s}, the main idea of this method
is to utilize the gravitational redshift, where the frequency of a photon is changed depending on
the gravitational potential. When the Earth is orbiting the Sun,
their distance varies by about 5.0 x 10$^{6}$~km during the year,
corresponding to 3.2 x 10$^{-10}$ in terms of relative frequency shift~\cite{kars}.
Hence, a precision level of about 0.1~ppb is needed to
be sensitive to an effect which could shift the Mu transition frequency with respect to
a frequency reference based on an ordinary matter system. The required sensitivity 
implies a 40-fold improvement
in the measurement of Mu 1S-2S transition frequency.

This improvement requires:
a) highest possible slow $\mu^{+}$ rate (e.g., 4000$\mu^{+}$/s at LEM~\cite{mue4,lemmusr} facility at PSI),
b) highest possible $\mu^{+} \rightarrow$ slow Mu conversion rate (e.g., Mesoporous silica~\cite{muinsio2}),
c) high power continuous wave laser (based on the development of, e.g., Ps 1S-2S at ETHZ~\cite{ps1s2s}) and
d) a better known reference frequency (e.g., I$_{2}$ calibration via frequency comb~\cite{mu1s2scomb}).

\subsection{Mach Zehnder interferometer}

\begin{figure}[htpb]
\centerline{
	\includegraphics[width=7cm]{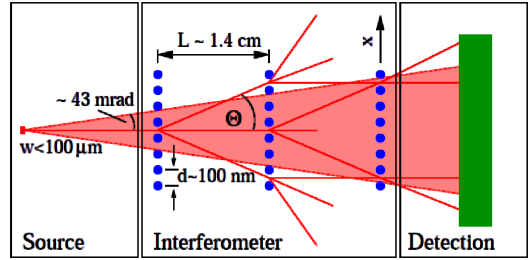}
}
\vspace*{8pt}
\caption{Mach Zehnder interferometer for muonium atom.\label{MZ}}
\end{figure}
Interferometry methods can be very sensitive and it was proposed to measure the 
phase shift of neutral antimatter caused by the Earth's gravitational field
in a transmission-grating interferometer~\cite{mzinter1,mzinter2,kirch,kaplan}
(see Fig.~\ref{MZ}). To apply this technique to Mu and in order to have enough 
statistics and a measurable deflection, one should aim for:
a) separation between gratings $\geqq$2.2~$\mu$s,
b) free standing grating pitch $\approx$100~nm and
c) $\geqq$10$^{5}$/s mono-energetic Mu beam (based on a superfluid helium source).

The precision of this method could reach $\frac{0.3\mathrm{g}}{\sqrt{\#\mathrm{days}}}$
and one could hope for 3\% from a 100 day measurement.

\section{Current status of our R\&D}

\subsection{1S-2S spectroscopy}
\subsubsection{Mu production with porous silica}

\begin{figure}[htpb]
\centerline{
	\includegraphics[width=5.0cm]{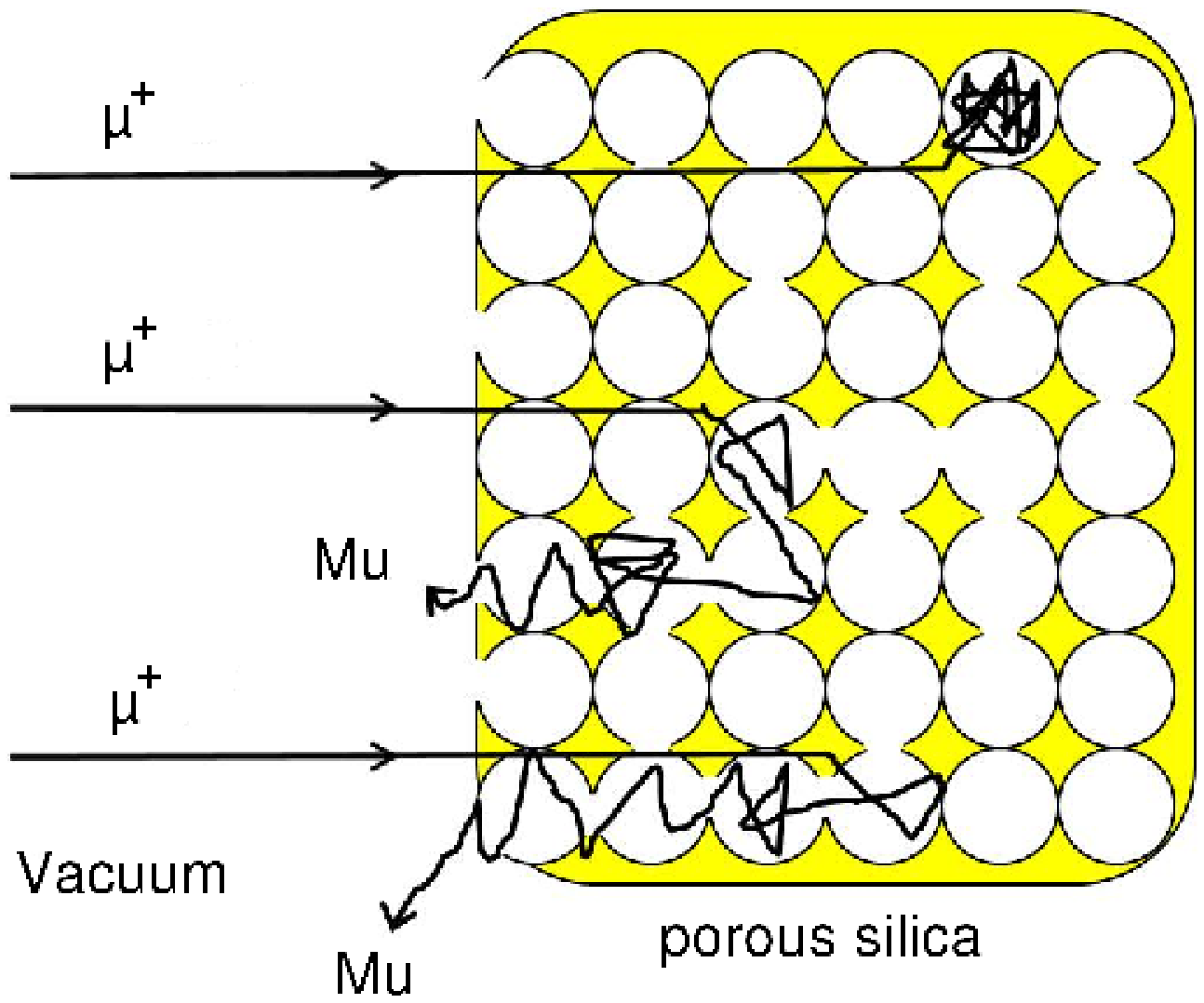}
	 \hfill
	\includegraphics[width=8.0cm]{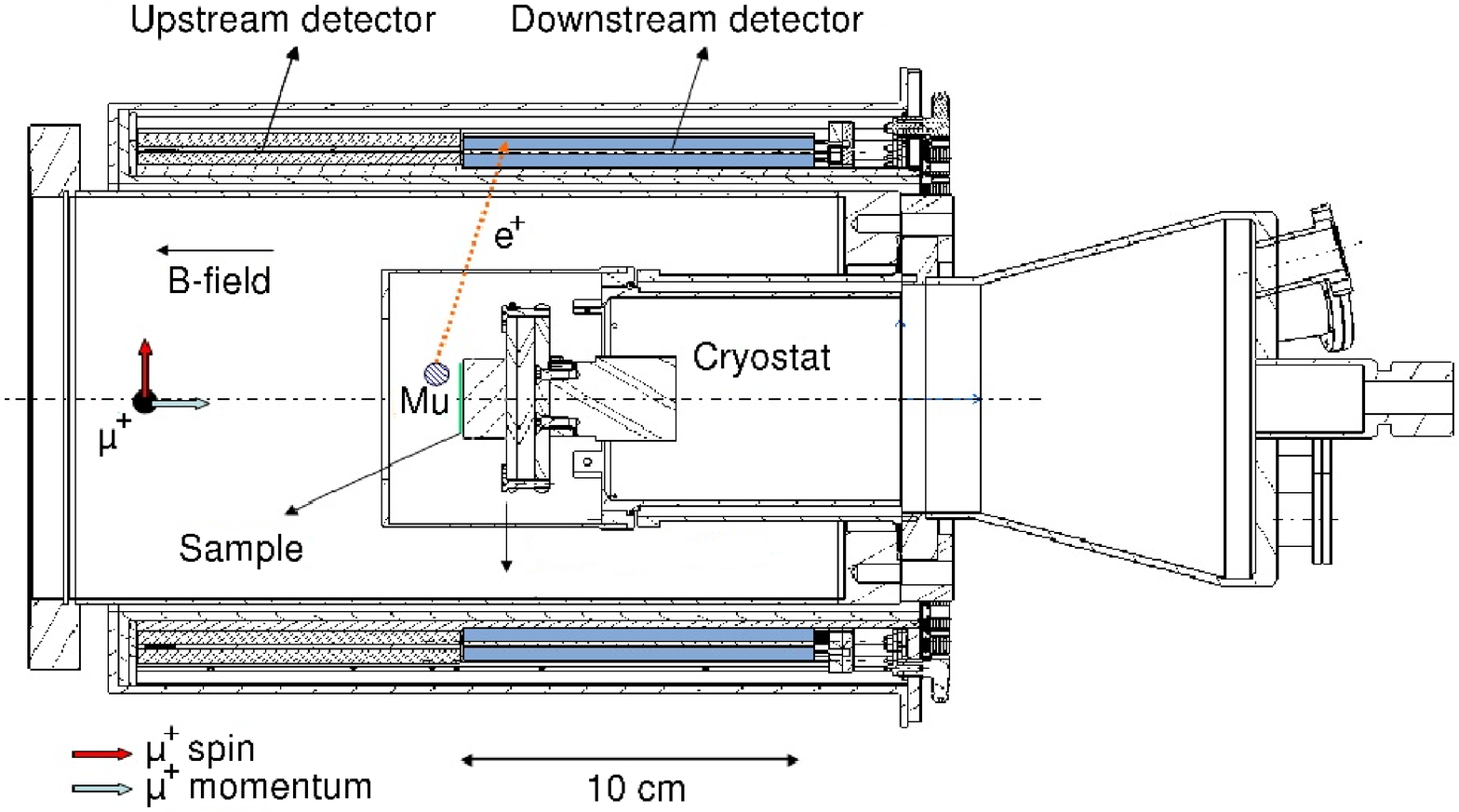}
}
\vspace*{8pt}
\caption{(left) Porous film of 1~$\mu$m thickness, a pore size of 5.0$\pm$0.5~nm, and a density of 1.1~g/cm$^{3}$.
(right) LEM sample chamber. The sample is glued on a silver coated
copper mount contacted to a cryostat. The sample is surrounded by
scintillators for positron detection grouped in upstream and
downstream counters. Each of them is additionally segmented in top,
bottom, left and right.\label{muproduction}}
\end{figure}

For high precision spectroscopy Mu in vacuum is needed to avoid matter effects.
Bulk silica has a high Mu formation rate but no emission from the target.
Silica with structured pore-networks (porous silica) could lead to a high fraction of Mu
diffusing out into vacuum (see Fig.~\ref{muproduction}).

Measurements have been carried out~\cite{muinsio2} at PSI's LEM with 4000/s $\mu^{+}$ on the sample. 
A muon spin rotation technique was used
to extract the Mu formation rate and a positron shielding technique allowed to extract 
the Mu yield in vacuum.

\subsubsection{Muon spin rotation technique ($\mu$SR)}

Due to parity violation of the weak interaction, decay positrons are emitted preferentially in direction
of $\mu^{+}$ spins. By monitoring the evolution of $\mu^{+}$ spins after implantation in an external
magnetic field, unbound $\mu^{+}$ and Mu can be distinguished because the Larmor precession frequency
of Mu is about 100 times larger than that of $\mu^{+}$.

With segmented positron detectors around the target, the Mu formation rate can be
determined from the disappearance of the $\mu^{+}$ precession signal.

\subsubsection{Positron shielding technique (PST)}

\begin{figure}[htpb]
\centerline{\includegraphics[width=10cm]{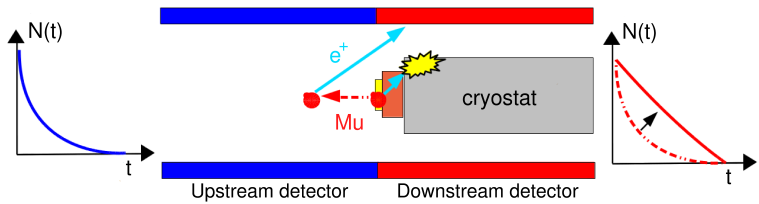}}
\vspace*{8pt}
\caption{A schematic illustration of positron shielding technique (PST).\label{pstmethod}}
\end{figure}

Without Mu emission into vacuum an exponential decay curve is observed with the downstream detector.
When there is Mu emission into vacuum, a deviation from the exponential curve appears 
due to less shielding when Mu decay outside of the sample (see Fig.~\ref{pstmethod}).

\begin{figure}[htpb]
\centerline{
\includegraphics[width=5.8cm]{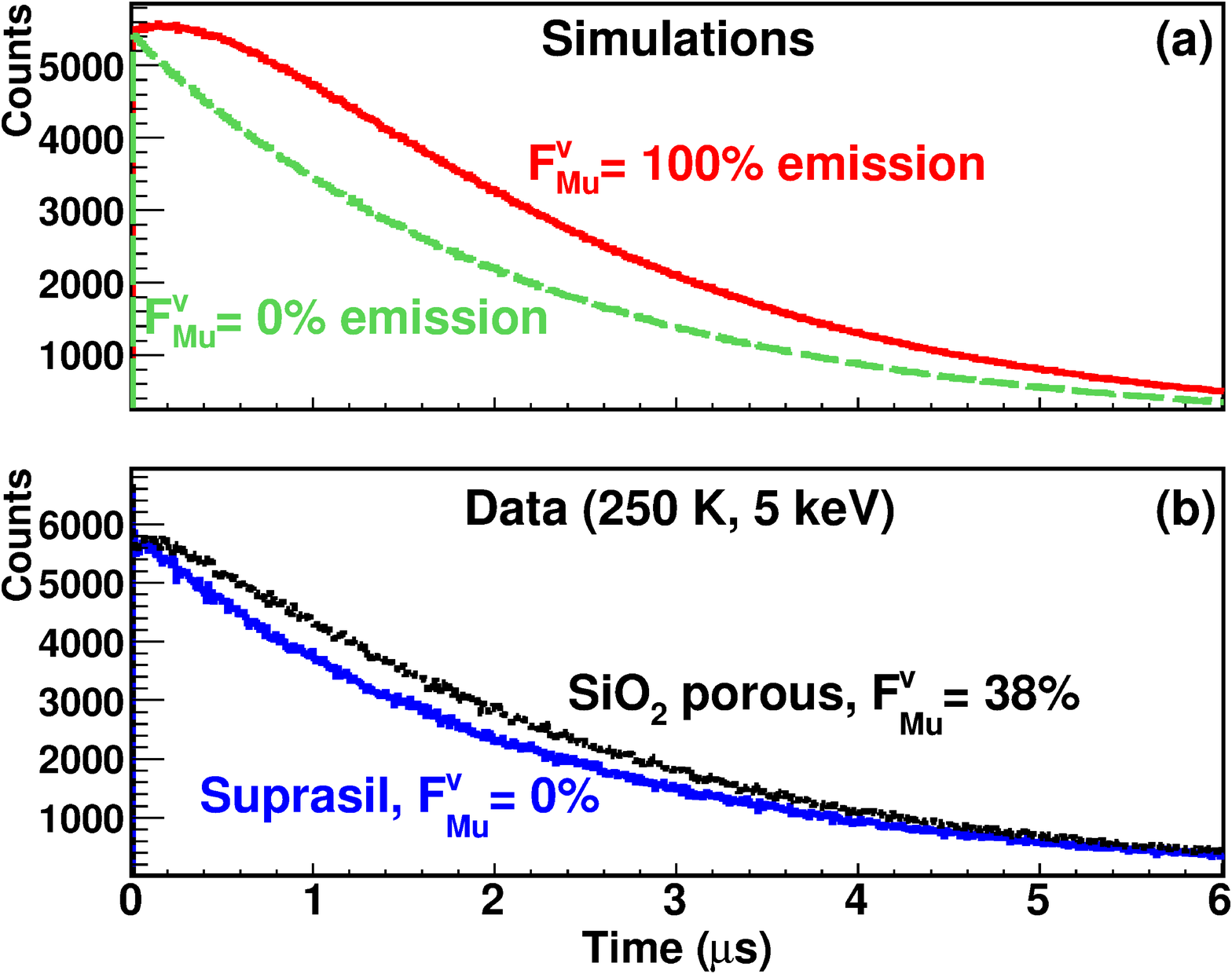}
\hfill
\includegraphics[width=7.4cm]{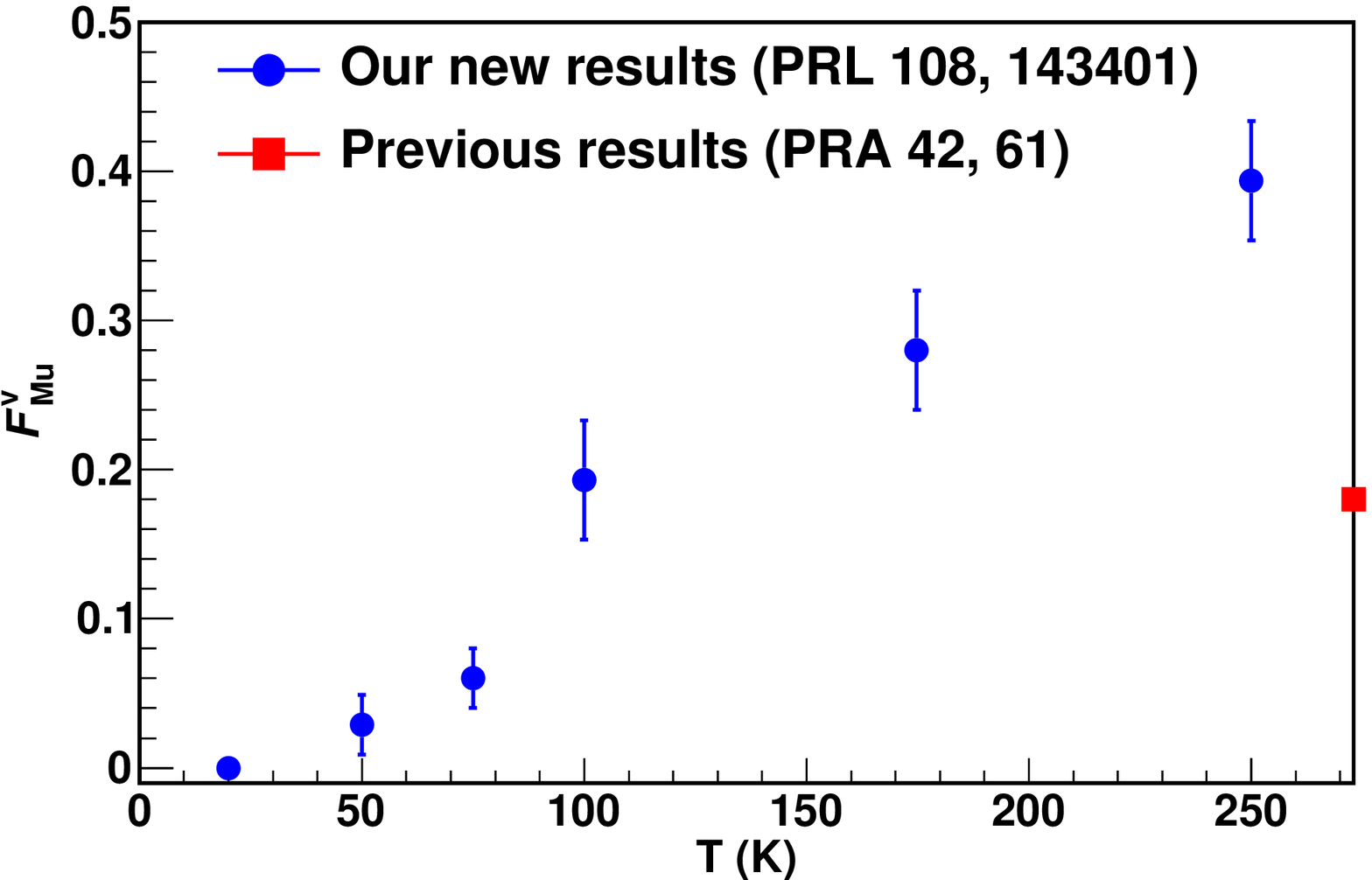}}
\vspace*{8pt}
\caption{(left) Simulated (a) and measured (b) positron spectra. A fit function
is used to extract the Mu yield in vacuum. 38\% was obtained for porous silica and 0\% was obtained
for suprasil (fused quartz) where zero Mu vacuum yield is expected. (right)
A temperature dependence of Mu vacuum yield for porous silica F-sample.\label{PSTspectra}}
\end{figure}

GEANT4~\cite{geant4} simulations were done for 0\% emission ($F_{0}$) and 100\% emission ($F_{100}$) and
the data were fitted with $F_{fit} = aF_{100} + (1-a)F_{0}$, where $a$ is the Mu
yield in vacuum. A typical result is shown in Fig.~\ref{PSTspectra}(left).

\subsubsection{Mu yield in vacuum}

We have studied the Mu yield in vacuum for different pore sizes and temperatures.
The best results obtained so far are 20\% at 100~K and 40\% at 250~K for 5~nm pore size (see Fig.~\ref{PSTspectra}).
There is evidence that a Boltzmann velocity distribution for Mu is preferred over a uniform velocity distribution. 

With these vacuum yields and available laser technology, it appears possible 
to improve precision in the Mu 1S-2S frequency by a factor of 10.

\subsection{Mu atom interferometry}
Before being able to apply a Mach-Zehnder interferometer to a Mu beam, such a beam must be developed. 
It will be based on a much improved slow muon beam and a novel Mu production and extraction to vacuum 
from superfluid helium.
\subsubsection{Slow muon beam}

\begin{figure}[htpb]
\centerline{\includegraphics[width=9cm]{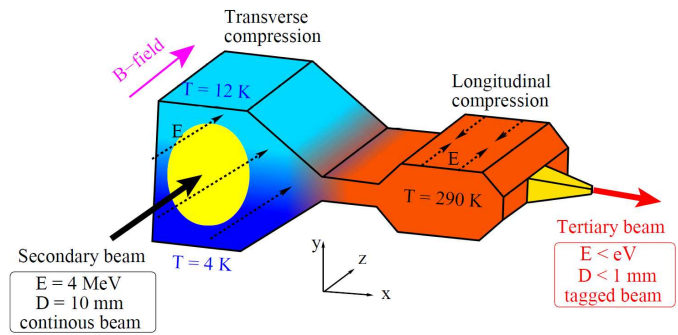}}
\caption{A schematic illustration of phase space compression.\label{PSC}}
\end{figure}

Phase space compression by a factor of $10^{10}$ compared to a standard
surface muon beam can be achieved by stopping muons in a few mbar of He gas, 
compressing the stop distribution and extracting
them back to vacuum~\cite{taqqu}. The compression uses a position-dependent $\mu^{+}$ drift in E and 
B fields in helium gas.
The final ultra slow $\mu^{+}$ beam with sub-mm size and sub-eV energy
can be re-accelerated (see Fig.~\ref{PSC}).

The compression scheme is divided into 3 stages: transverse compression, longitudinal compression
and extraction of the $\mu^{+}$ beam. The efficiency of the scheme is estimated to about 0.1\%, 
mainly limited by the muon lifetime.

\subsubsection{Longitudinal compression}

\begin{figure}[htpb]
\centerline{\includegraphics[width=7.0cm]{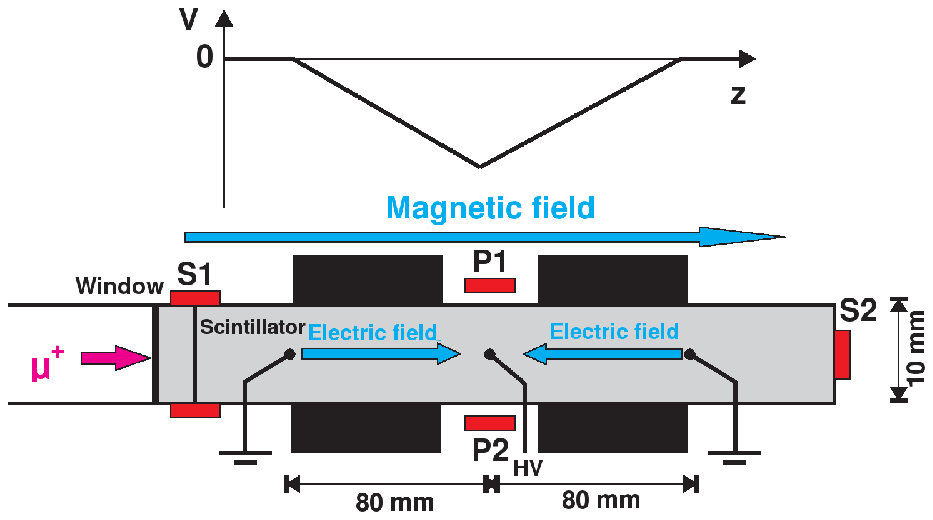}
\includegraphics[width=7.2cm]{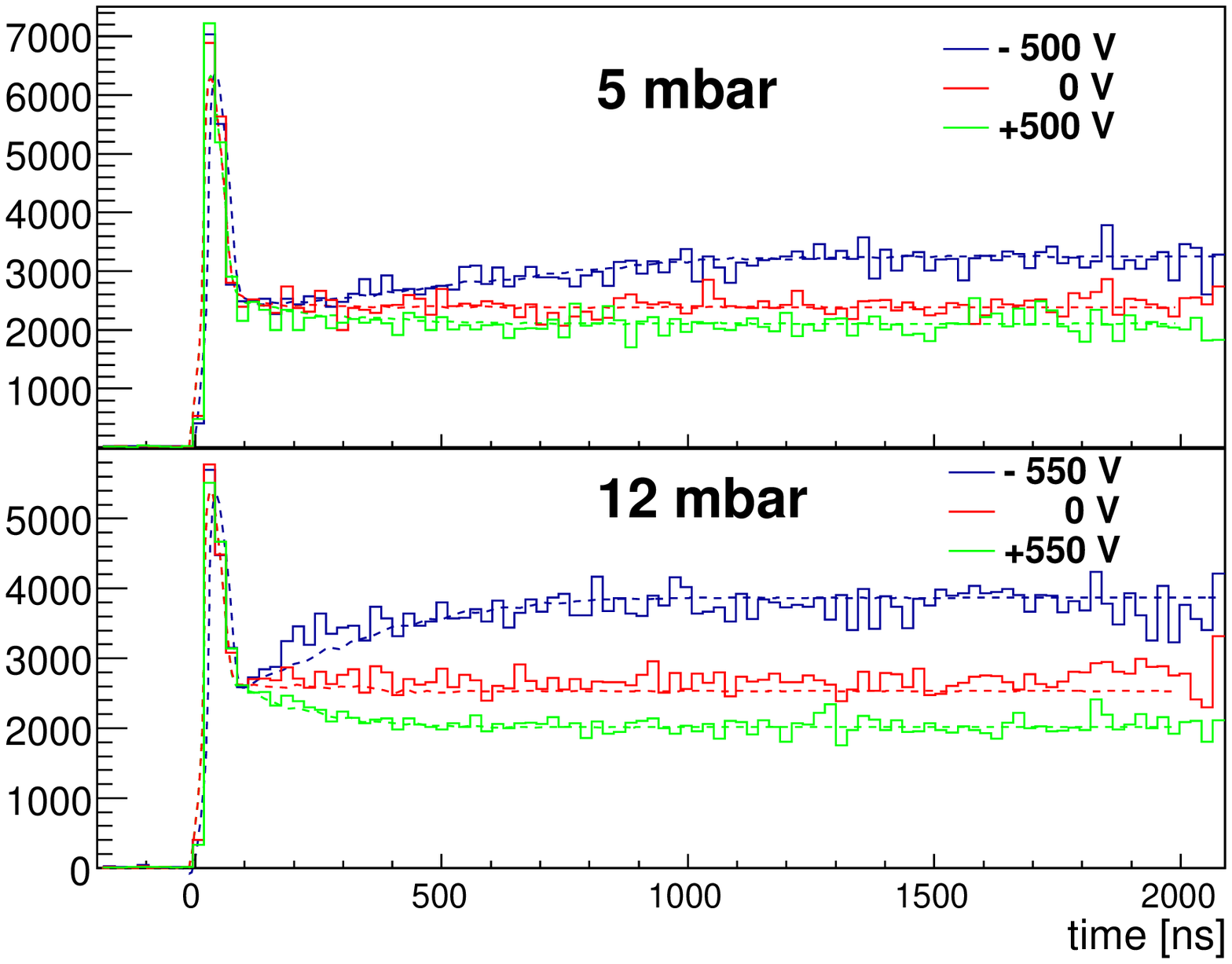}
}
\vspace*{-8pt}
\caption{(left) A schematic illustration of longitudinal compression.
	(right) Measured (continuous lines) and simulated (dotted) positron spectra divided by $e^{-t/2200}$
	for (+550~V,0~V,-550~V) electric fields at 5~mbar and 12~mbar helium gas pressure.
\label{mucool}}
\end{figure}

Longitudinal compression was demonstrated at PSI's $\pi$E1 beam line using $\mu^{+}$ of 10~MeV/c at a
rate of 2 x 10$^{4}$/s (see Fig.~\ref{mucool}). Low energy $\mu^{+}$ elastic collision physics 
and Mu formation in helium gas, which were scaled from available data for protons~\cite{scaling,elastic}, were implemented into GEANT4, and simulations were done to compare with the data.
In Fig.~\ref{mucool} (right), time spectra of detected positrons at P1 and
P2 are displayed after a $\mu^{+}$ trigger in S1. To eliminate lifetime effects,
the spectra were divided by $e^{-t/\tau_{\mu}}$ transforming an exponential decay
to a uniform distribution. When $\mu^{+}$ are drifting to the central
region, the detection efficiency and count rate increase, as is seen for negative voltage.
The opposite effect occurs when the polarization of the electric field is reversed.
Good agreement between simulation and data was achieved and compression of the 16~cm wide muon
swarm into 0.5~cm width occurs in much less than 2~$\mu$s. This result shows that the compression
process is faster than the mean lifetime of $\mu^{+}$ and hence the longitudinal compression
is feasible~\cite{yubao}.

\subsubsection{Helium gas density gradient}

One challenge of the transverse compression part is to realize a helium gas density gradient
over the muon stop distribution via a static temperature gradient. Turbulence must
be avoided by operating the lower side of the target at lower temperature.

\begin{figure}[htpb]
\centerline{\includegraphics[width=7cm]{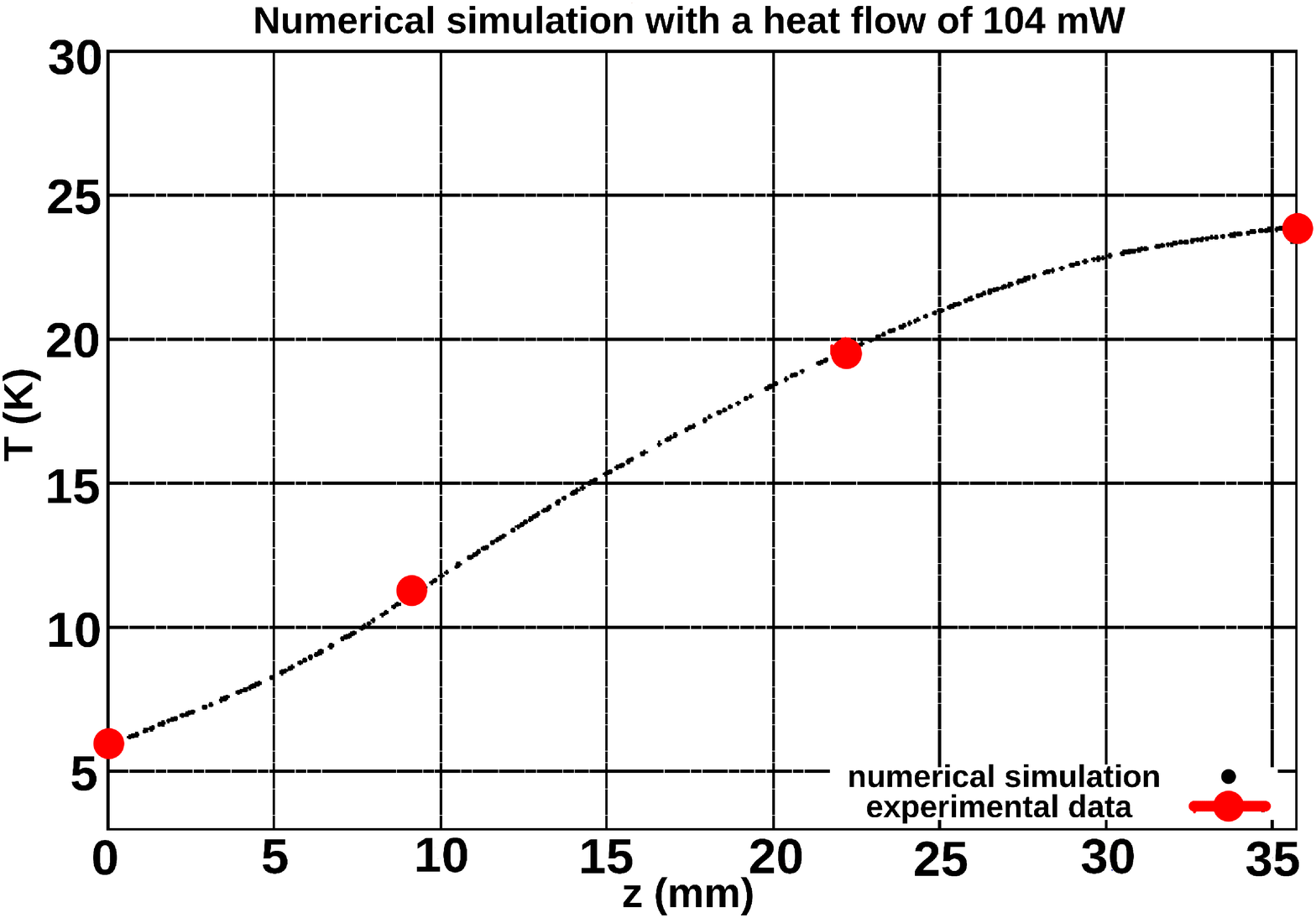}}
\vspace*{8pt}
\caption{Comparison of simulation and measurements of the gas cell with density gradient.\label{gascellresult}}
\end{figure}

We have made a first demonstration of a stationary temperature gradient in the laboratory. 
A cylinder made of copper (top, bottom) and thin stainless steel (sides) was attached to
a cryostat and equipped with thermometry and heaters. The cylinder was filled with
helium pressures of 0.01~mbar to 50~mbar and stationary temperatures have been measured and
simulated. Good agreement between a COMSOL
simulation and measured data was obtained~\cite{gunther} (see Fig.~\ref{gascellresult} for an example)
proving the feasibility for a muon target assembly.

\section{Conclusion and Outlook}

So far our feasibility studies into Mu production for 1S-2S spectroscopy and for atom interferometry
yield very promising results. High Mu vacuum yields have been demonstrated for suitable porous silica
targets opening up the path for a next generation spectroscopy experiment. 
Towards a high quality slow $\mu^+$ beam, longitudinal compression has been demonstrated and
the feasibility for a gas target with helium density gradient was shown. The next step for the beam development
will be the demonstration of transverse compression of a muon stop distribution. In a separate experiment
we aim at remeasuring the Mu production in superfluid helium below 0.5~K~\cite{muinhe} and verify the
predicted quasi-monoenergetic emission into vacuum~\cite{mufromhe,taqqu2}. Studies of a 
Mach-Zehnder interferometer will be pursued in collaboration with IIT~\cite{kaplan}.

\section*{Acknowledgments}

This work was performed at the ETH Zurich and at the Paul Scherrer Institut. We gratefully acknowledge the support of the
accelerator and beamline groups. Our work greatly benefitted from using the $\mu$E4 LEM installation and the new $\pi$E1 area
and intense collaboration with Thomas Prokscha, Konrad Deiters and Claude Petitjean, the ETH group ‘Precision Physics at Low Energy’
and the PSI ‘Muon Physics Group’. We would like to thank Aldo Antognini, Paolo Crivelli, Dan Kaplan, Tom Phillips, Florian Piegsa,
and David Taqqu for fruitful discussions in connection with this conference presentation.
This work was supported by the Swiss National Science Foundation grants
200021-129600 and 200021-146902.

\end{document}